\documentclass[journal]{IEEEtran}  
\usepackage{cite}
\usepackage{amsmath,amssymb,amsfonts}
\usepackage{algorithmic}
\usepackage{graphicx}
\usepackage{textcomp}
\usepackage{booktabs}
\usepackage{array}
\usepackage{xcolor}
\usepackage{placeins}
\usepackage{algorithm}
\usepackage{float}
\usepackage{url}
\usepackage{capt-of}
\DeclareMathOperator*{\argmin}{arg\,min}

\begin{document}

\title{Benchmark of Bayesian Optimization and Metaheuristics for Control Engineering Tuning Problems with Crash Constraints}

\author{David Stenger\textsuperscript{1} and Dirk Abel\textsuperscript{1}
	\thanks{\textsuperscript{1}Institute of Automatic Control (IRT), RWTH Aachen University, Germany, {\tt D.Stenger@irt.rwth-aachen.de}}
}
\maketitle

\begin{abstract}
	Controller tuning based on black-box optimization 
	allows to automatically tune performance-critical parameters w.r.t. mostly arbitrary high-level closed-loop control objectives. 
	However, a comprehensive benchmark 
	of different black-box optimizers for control engineering problems has not yet been conducted. Therefore, in this contribution, 11 different versions of Bayesian optimization (BO) 
	are compared with seven metaheuristics and other baselines 
	on a  
	set of ten  
	deterministic simulative single-objective tuning problems in control.    
	Results indicate that deterministic noise, low multimodality, and substantial areas with infeasible parametrizations (crash constraints) characterize control engineering tuning problems.
	Therefore, a 
	flexible method to handle crash constraints with BO is presented. A resulting increase in sample efficiency is shown in comparison to standard BO. Furthermore, benchmark results indicate that pattern search (PS) performs best on a budget of $\mathbf{25 \, d}$ objective function evaluations and a problem dimensionality $\mathbf{d}$ of $\mathbf{d = 2}$. Bayesian adaptive direct search, a combination of BO and PS, is shown to be most sample efficient for  $\mathbf{3 \le d \le 5}$. Using these optimizers instead of random search increases controller performance by on average $\mathbf{6.6 \%}$ and up to $\mathbf{16.1 \%}$.  
\end{abstract}

\begin{IEEEkeywords} 
	Automatic Controller Tuning, Bayesian Optimization, Metaheuristics, Crash Constraints
\end{IEEEkeywords}

\section{Introduction}
\label{sec:introduction}

Algorithms used in control engineering for e.g. control, state estimation or planning often rely on a number of performance critical tuning parameters.  
Examples include controller gains $k_\mathrm{p}$ and $k_\mathrm{i}$ of PI-controllers, and  
the weighting matrices 
in case of model predictive control (MPC) and Kalman filters. An important challenge in applying these algorithms  
is to set the tuning parameters in an optimal way.  

Analytical or empirical tuning laws 
only exist for a limited number of combinations of plants, algorithms and performance criteria (e.g. the LQR method for linear systems, state feedback controllers, and quadratic objective functions). Manual tuning  
on the other hand can 
potentially be tedious and suboptimal.

A promising alternative is to formulate the parameter tuning problem as an episodic black-box optimization problem. 
Using this approach, the closed-loop performance 
is iteratively evaluated on repetitive simulative or experimental episodes with different parameters. An optimizer chooses the parameters in order to directly optimize  
arbitrary  
high-level control objectives such as energy consumption, product quality, and comfort.

Objective function evaluations may be expensive e.g. in terms of CPU time 
in case of high fidelity plant models.  
As a result, the sample efficiency, i.e. the ability of optimizers to find good solutions with as little objective function evaluations as possible, is of key importance.
An additional challenge is posed by 
unsuccessful simulations or experiments, where no useful objective function value can be obtained. E.g., the closed-loop may become unstable for some unsuitable parameter combinations.     
This setting
is known as \textit{learning with crash constraints (LCC)} \cite{Marco.2021}.    

Historically, metaheuristics, e.g. 
particle swarm optimization (PSO), were popular choices for approximately solving these tuning problems.  
In recent years, Bayesian optimization (BO) has become popular.
LCC poses a challenge specifically to BO. Therefore, a heuristic BO extension using virtual data points 
is presented. In contrast to literature, it does not require domain knowledge or additional models and therefore can flexibly be incorporated within other BO-extensions i.e. contextual or constrained BO.

BO is often claimed to be more sample efficient than metaheuristics. 
However, its sample efficiency has not yet been  
compared to other optimizers on a wide variety of tuning problems in control engineering. The same is true for the impact of the various algorithmic design choices 
within BO. 
Comprehensive benchmarks 
were  
conducted on synthetic benchmarks (e.g. \cite{LeRiche.2021}), 
in other domains such as machine learning (e.g. \cite{Snoek.2012}) or for specific applications and algorithms (e.g. \cite{Calandra.2016, NeumannBrosig.2019}).  
In contrast, in this study, we 
compare the sample efficiency of 11 BO variants with seven other optimizers on ten deterministic unconstrained single-objective simulative tuning problems. 

The main \textbf{contributions} of this article are: 

\begin{itemize}
	\item A flexible BO method for LCC using virtual data points
	\item Characterization of the objective function landscapes of ten different controller tuning problems 
	\item A comparison of different versions of BO  
	with meta-heuristics 
	and other benchmarks 
	in terms of sample efficiency 
	\item Analysis of the practical relevance of optimizer choice on time-domain performance
\end{itemize}
 
The paper is structured as follows: Sec. \ref{sec:Problem} states the problem. 
Sec. \ref{sec:relwork} introduces related work. 
Afterwards, in Sec. \ref{sec:BO}, BO 
including the novel LCC method 
is introduced. Sec. \ref{sec:Benchmark} 
analyzes the  
benchmark test cases. Benchmark results   
are presented (cf. Sec. \ref{sec:res}) and discussed (cf. Sec. \ref{sec:Dis}). Concluding remarks are given in Sec. \ref{sec:con}.

\section{Problem Statement} \label{sec:Problem}

\subsection{Optimization Problem Formulation}

\begin{figure}[h]
	\centering
	\includegraphics[width=0.8\linewidth]{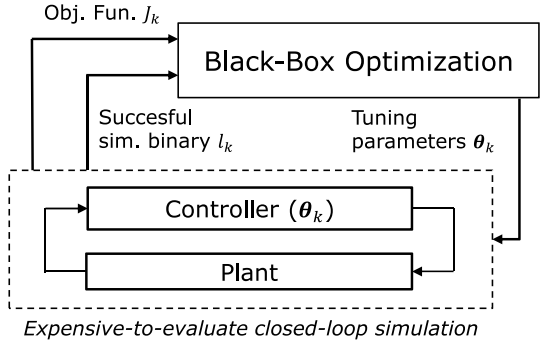}
	\caption{Considered black-box optimization setting.}
	\label{fig:BBOverview}
\end{figure}

The black-box optimization setting considered in this contribution is depicted in Fig. \ref{fig:BBOverview}. A closed-loop simulation consisting of a controller and/or observer, as well as a plant model is used. This simulation is sequentially queried with different tuning parameter combinations $\boldsymbol{\theta}_k$. From the time-domain response of the closed-loop system, the objective function value $J_k$ as well as a binary $l_k$, indicating whether the simulation was successful or not, is calculated and returned to the black-box optimization.   

The black-box optimizer has the task of finding optimal parameters $\boldsymbol{\theta}^*$ to approximately solve the problem

\begin{equation}
	\begin{aligned} \label{eq:BO_OptProb}
		\boldsymbol{\theta}^* =  \argmin_{\boldsymbol{\theta} \in \mathbb{R}^d} \qquad &  J(\boldsymbol{\theta}) \\
		\mathrm{s.t. }\qquad \qquad \qquad \qquad  & \boldsymbol{\theta}_{\mathrm{min}} \le \boldsymbol{\theta} \le \boldsymbol{\theta}_{\mathrm{max}}\\
		& l(\boldsymbol{\theta}) = 1 \, ,
	\end{aligned}
\end{equation}

with the objective function $J(\boldsymbol{\theta})$.
Due to the simulation's black-box nature, gradients are not available. The tuning parameters $\boldsymbol{\theta}$ are subject to box constraints indicated by $\boldsymbol{\theta}_{\mathrm{min}}$ and $\boldsymbol{\theta}_{\mathrm{max}}$. 

Additionally, a so-called \textit{crash constraint} \cite{Marco.2021} $l(\boldsymbol{\theta}) = 1$ applies. If poor parameters lead to e.g. a diverging filter solution or unstable closed-loop behavior, 
the simulation may crash or be aborted. A value of $l(\boldsymbol{\theta}) = 0$ indicates that no sensible (e.g. extremely large) value for the objective function is available (see also Sec. \ref{sec:TestCases}).

\subsection{Design Choices in BO with Crash Constraints} \label{sec:DCBO}

\begin{figure}[th]
	\centering
	\includegraphics[width = 0.45 \textwidth]{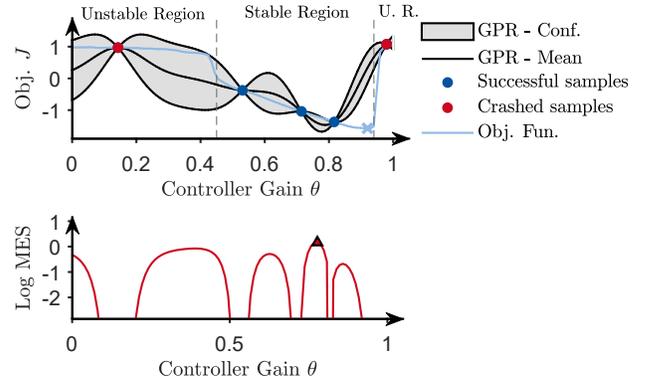}
	\caption{BO example without crash constraint handling. \textbf{Top}: Objective function, evaluations, and GP model. \textbf{Bottom}: Acquisition function.}
	\label{fig:wocrash}
\end{figure}

Fig. \ref{fig:wocrash} visualizes the basic idea of BO. The goal is to use all information obtained so far through previously objective function evaluations, to find the most promising next sample. To achieve that, a probabilistic black-box model, in this case Gaussian Process Regression (GPR), with mean  
and predicted uncertainty  
is fitted to past objective function evaluations. 
This model is used by an acquisition function (Fig. \ref{fig:wocrash}, bottom), in this case max-value entropy search (MES), to determine the utility of sampling at a given location. Maximizing the acquisition function determines the next sample point (red triangle). The next sample will subsequently be evaluated on the expensive-to-evaluate black-box simulation to obtain the corresponding objective function value. The unknown objective function with its optimum (x) is displayed in light blue. 

The main challenge associated with crash constraints is also highlighted in Fig. 2. At the border between stable and unstable regions, the objective function may exhibit steep gradients or discontinuities. This  
contradicts the smoothness assumptions encoded in GPR, and therefore deteriorates BO performance. In this example, the objective function is overestimated, and therefore, the acquisition function is rather small at the location of the global optimum. 

The main design choices of BO, which will be examined in this contribution (cf. Sec. \ref{sec:BO}), are:  
\begin{itemize}
	\item the structure of the GPR, namely prior mean and kernel, 
	\item the acquisition function,
	\item and how to address crash constraints.
\end{itemize}

\section{Related work} \label{sec:relwork}

\subsection{Bayesian Optimization in Control Engineering } \label{sec:ControllerTuning}

BO for controller tuning can be attributed to the \textit{micro Data RL} branch of machine learning (ML) \cite{Chatzilygeroudis.2020} and is sometimes termed data efficient policy search.
It has been widely applied 
to automatically tune the hyperparameters of  various different algorithms.  Examples include 
model predictive control (MPC) \cite{Andersson.16.05.201621.05.2016, Stenger.2020, GharibAli.2021}, 
LQR \cite{Calandra.2016, Marco.2016}, 
PID \cite{Chen.2019, Fiducioso.6282019, Khosravi.2019},
 and Kalman filter \cite{Gehrt.1123202011242020, Stenger.2022, Chen.71020187132018, Riva.2022}. 
Areas of application  are wide spread including autonomous driving \cite{Andersson.16.05.201621.05.2016, GharibAli.2021} and  robotics \cite{Marco.2016, Calandra.2016, Stenger.2022}. 
It was also shown that BO can be applied to 
complex hierarchical controller structures 
\cite{MohammadKhosravi., Roveda.2020, Stenger.2022}, 
and be used to fairly compare different algorithms \cite{M.Manhaes.2017}. 
Promising experimental results were reported e.g. in  \cite{Frohlich.06.10.2021, NeumannBrosig.2019, GianlucaSavaia.2021, Konig.19.01.2021, Konig.28.10.2020, Marco.2016, Roveda.2020, Berkenkamp.14.02.2016, Calandra.2016}.

BO can also address various  
advanced challenges in controller tuning such as 
optimization with unknown constraints (e.g. \cite{Stenger.2020, Stenger.2022, Konig.28.10.2020}).  
In Safe BO, the goal is to prevent the optimizer from sampling parameters potentially leading to unsafe (e.g. unstable) behavior. 
Safe areas of the parameter space can either be learned \cite{Berkenkamp.14.02.2016, Konig.19.01.2021, Fiducioso.6282019} or  
deduced from models \cite{Dorschel.2021}. Additionally, information from simulation 
and experiments can be combined \cite{Marco.2017} and parameters optimized as a function of different contexts \cite{Fiducioso.6282019}.  
Multi-objective BO allows to search for the pareto front of different conflicting objectives \cite{GharibAli.2021} and research has been done on controller specific kernels \cite{MarcoAlonso.2017}, preference-based tuning \cite{ZhuMengjia.2021}, robust optimization \cite{Stenger.2020, Stenger.2022, Frohlich.2020}, and time-varying objective functions \cite{Brunzema.2022}. 

The presented applications and extensions show that BO is a 
flexible tool for automatic tuning in control. 
However, to 
the authors' knowledge, a comprehensive benchmark study of the various design choices in BO and a comparison to other optimizers was not done for a representative amount of control engineering problems.     
Often different settings of a developed algorithms are proposed or compared to algorithms designed for the considered specific problem class such as safe BO  \cite{Andersson.16.05.201621.05.2016, Stenger.2020, GharibAli.2021, Fiducioso.6282019, Khosravi.2019, Konig.19.01.2021, Konig.28.10.2020, Berkenkamp.14.02.2016}. Different acquisition functions have been compared for one specific application e.g. in \cite{Calandra.2016, NeumannBrosig.2019}.
The sample-efficiency of BO compared to baselines e.g. PSO is also only examined for single applications e.g. \cite{Calandra.2016, Dorschel.2021}

\subsection{Bayesian Optimization with Crash Constraints} \label{sec:rwLCC}

Learning with crash constraints (LCC) \cite{Marco.2021} deals with situations where objective function values are not available or not sensible (e.g. they grow extremely large) for specific parametrizations. 
We refer to \cite{Marco.2021} for a comprehensive literature review.  
Heuristic approaches include the usage of 
a probabilistic classifier in combination with constrained BO (e.g. \cite{DavidV.Lindberg.2015, Stenger.2020}). 
Alternatively, a fixed penalty can be assigned (e.g. \cite{Marco.2016}) or data, obtained before the crash, can be used (e.g. \cite{Calandra.2016}). However, it may require substantial domain knowledge in order
to design the objective function such that  smoothness at the borders between crashed and successful evaluations is preserved. 
In \cite{Marco.2021}, a non-heuristic method is introduced and evaluated experimentally. A GP model capable of combining regression and classification is combined with constrained BO.  

In contrast to the literature, the method presented in Sec. \ref{sec:crashConstraints} does not require domain knowledge or the training of additional models and is therefore easily usable in conjunction with other BO extensions. The method was superficially introduced and applied to one constrained hierarchical controller tuning problem in our previous conference paper \cite{Stenger.2022}.

\subsection{Metaheuristics for Controller Tuning} \label{sec:ControllerTuningMeta}

In addition to BO, metaheuristics are also widely used for controller tuning. For example in \cite{SidharthaPanda.2008, Artale.2017, OuChao.2006} it was observed that PSO is more sample efficient than GA for one specific tuning task each. 
In \cite{SinghMahesh.2016} pattern search, simulated annealing, GA and PSO were compared on two PID controller optimization tasks. 
Pattern search was found to be competitive although substantially less objective function evaluations were required.
Various other metaheuristics e.g. firefly algorithm (FA), PSO, ant colony optimization (ACO), bat algorithm (BA) and imperialist competitive algorithm (ICA) were also compared in \cite{KusumaDwiHendra.2016}. 
A combination of a metaheuristics and BO has been used for controller tuning in \cite{RikkyR.P.R.Duivenvoorden.2017}, however no comparison to only using BO or PSO was given.

In comparison to the BO literature, larger budgets are typically used and the focus is placed on tuning in simulation. Although, typically different metaheuristics are compared in the studies above, only single applications are considered. 

\subsection{Optimization Benchmarks in Other Domains}

In \cite{LeRiche.2021}, various variants of BO with other sample efficient optimizers are compared on the synthetic COmparing Continuous Optimizers (COCO)\cite{NikolausHansen.2021} benchmark. 
Results indicate that BO excels in small dimensions $d \le 5$ and a budget of $10 \, d$ to $20 \, d$. It was found that the Mat\'{e}rn 5/2 (MA) kernel may be surpassed by the squared exponential (SE) kernel on some problems \cite{LeRiche.2021}. Also, using a quadratic instead of a constant prior mean may be beneficial in some cases \cite{LeRiche.2021}. This is in agreement to e.g. \cite{Stenger.2019}, where automatic prior mean selection improved performance.

In the ML domain, Bayesian adaptive direct search (BADS), a hybrid method of PS and BO was compared with other optimizers, e.g. BO, PSO, GA etc. on various real world model fitting problems \cite{Acerbi.2017}. BADS was shown to outperform the other competitors, 
highlighting the potential of hybrid methods. 
In \cite{Snoek.2012}, different variants of BO were compared on practical ML hyperparameter optimization problems. It was shown that 
the MA kernel outperforms the SE kernel.
Furthermore, in \cite{Turner.2021} 
it was shown that BO consistently outperformed random search in a ML hyperparameter optimization competition. Top competitors used BO ensembles making use of various acquisition functions and surrogate models, indicating that different BO setups are best for different problem settings. Differential evolution (DE) was also used in some of the ensembles. 

\section{Bayesian Optimization with Crash Constraints} \label{sec:BO}

BO was briefly introduced visually in Sec. \ref{sec:DCBO}. Here we focus on the implementation details, the evaluated algorithmic settings, and the novel LCC method. A further in-depth introduction to BO can be found for example in \cite{Shahriari.2016, garnett_bayesoptbook_2022}. Algorithm \ref{Algo:BayesOpt} gives an overview over BO with crash constraints.  

\begin{algorithm}[h] 
	1: Initial sampling of $\Theta_{1}$, $\mathcal{J}_{1}$ and $\mathcal{L}_{1}$:  \\ [3pt]
	2: \textbf{for} k = 1; 2; . . . ; \textbf{do} \\[3pt]
	3: \quad $\hat{J}_{k} \leftarrow \textrm{addVirtualData}(\Theta_{k},\mathcal{J}_{k},\mathcal{L}_{k}$) (Sec. \ref{sec:crashConstraints}) \\[3pt]
	4: \quad update GPR surrogate model using  
	$\Theta_{k}$, $\mathcal{J}_{k}$ and $\hat{\mathcal{J}}_{k}$\\[3pt]
	5: \quad select $\boldsymbol{\theta}_{k+1}$ by optimizing an acquisition function:\\ 
	\hspace*{6.5mm} $\boldsymbol{\theta}_{k+1}^\prime = \argmin_{\boldsymbol{\theta}} \quad \alpha(\boldsymbol{\theta}|\Theta_{k},\mathcal{J}_{k} , \hat{\mathcal{J}}_{k},)$\\[3pt]	
	6: \quad query objective function to obtain $J_{k+1}^\prime$ and $l_{k+1}^\prime$ \\[3pt]
	7: \quad augment data: $\Theta_{k+1} = [ \Theta_{k}, \boldsymbol{ \theta}_{k+1}^\prime]$,\\ \hspace*{6.5mm} $\mathcal{J}_{k+1} = [ \mathcal{J}_{k},\mathcal{J}_{k+1}^\prime]$,  
	  $\mathcal{L}_{k+1} = [\mathcal{L}_{k},l_{k+1}^\prime]$ \\[3pt]
	8: \textbf{end for} 
	\caption{Bayesian Optimization with Crash Constraints}
	\label{Algo:BayesOpt}
\end{algorithm} 
First, an initial sampling is conducted to obtain an initial set of evaluated parameters $\Theta_{1}$, objective function values $\mathcal{J}_{1}$, and crash binaries $\mathcal{L}_{1}$ (cf. Step 1). Here, we use $d + 1$ randomly selected initial samples. Afterwards, at each iteration $k$, virtual objective function values are calculated adaptively for all crashed evaluations to obtain the augmented objective function value set $\hat{\mathcal{J}}_{k}$ (cf. Step 3 and Sec. \ref{sec:crashConstraints}).         
GPR \cite{CarlEdwardRasmussenandChristopherK.I.Williams.2006} is used as a surrogate model in order to approximate the true unknown objective function landscape in a probabilistic manner (cf. Step 4). Several different settings for the GPR model are evaluated in this benchmark paper (cf. Sec. \ref{sec:GPRMdl}).     
The GPR model is used within an acquisition function $\alpha$ to determine the utility of the evaluation of a given parametrization for the progress of the optimization. The acquisition function is maximized to obtain the next sample $\boldsymbol{ \theta}_{k+1}^\prime$ (cf. Step 5). In this contribution, three different types of acquisition functions are compared (cf. Sec. \ref{sec:AccMdls}). The next sample is then evaluated on the expensive-to-evaluate objective function obtaining $J_{k+1}^\prime$ and $l_{k+1}^\prime$ (cf. Step 6). Afterwards the data set is augmented with the obtained black-box responses (cf. Step 7) and the next iteration is reached.

\subsection{Gaussian Process Surrogate Model} \label{sec:GPRMdl}

GPR yields a Gaussian distribution $\tilde{J} \left( \boldsymbol{\theta} \right)$ with mean $\mu_J \left( \boldsymbol{\theta} \right)$ and standard deviation $\sigma_J \left( \boldsymbol{\theta} \right)$ for the objective function as a function of parameters $\boldsymbol{\theta}$ and data $\mathcal{D}$ :
\begin{equation}
	\tilde{J} \left( \boldsymbol{\theta} | \mathcal{D} = \{ \Theta_{k}, J_k \} \right) \sim \mathcal{N} \left(\mu_J \left( \boldsymbol{\theta} \right), \sigma^2_J \left( \boldsymbol{\theta} \right)\right) \, .
\end{equation}
In order to do so, the objective function values $J_i$ at locations $\boldsymbol{\theta}_i$ are modeled as a Gaussian process (GP): $(J_i) \sim \mathrm{GP} \left(m,k \right)$. 
The GP can be seen as a probability distribution over possible objective functions. Since test cases are deterministic,
no observation noise is present\footnote{In order to avoid numerical difficulties when inverting the covariance matrix, a constant regularization term of $\sigma_n = 4.5400e-05$ is added to the main diagonal of the covariance matrix. This resembles a slight artificial observation noise.}.   

The two main degrees of freedom in GP models are the kernel and prior mean function. A kernel function with hyperparameters $\theta_{\mathrm{GPR,K}}$ 
determines the correlation of the objective function values $J_{i}$, $J_{j}$ as a function of their respective locations in input space $\boldsymbol{\theta}_i$, $\boldsymbol{\theta}_j$: $\mathrm{Cov}\left(J_{i},J_{j} \right) =k\left(\boldsymbol{\theta}_i,\boldsymbol{\theta}_j|\theta_{\mathrm{GPR,K}}\right)$.
Here we compare the squared exponential (SE) kernel with the Mat\'{e}rn 5/2 (MA) kernel. In both cases automated relevance determination (ARD), i.e, different kernel length scales for each dimension, is used. The SE kernel gives higher probability to very smooth (infinitely differentiable) functions, whereas the MA kernel favors less smooth functions. As a result, MA is usually the default kernel choice in BO toolboxes \cite{LeRiche.2021}. However, is some cases the SE kernel was shown to be beneficial (cf. Sec. \ref{sec:relwork}).

Secondly, the prior mean function with hyperparameters $\theta_{\mathrm{GPR,m}}$, determines the trend of the GP model in absence of nearby data: $\mathrm{E} \left[J_{i}\right] = m(i) = m(\boldsymbol{\theta_i} \mid \theta_{\mathrm{GPR,m}})$. 
Usually a constant mean is used but a quadratic mean may be beneficial in some cases (cf. Sec. \ref{sec:relwork}). Therefore, we compare both cases.

GPR hyperparameters $\theta_{\mathrm{GPR,K}}$, $\theta_{\mathrm{GPR,m}}$ parameterize the mean and covariance functions. They are optimized at each model update step (cf. Algo. 1: Step 4 ) by maximizing their posterior probability. A combination of random search and gradient-based optimization is used for hyperparameter optimization to deal with the possibly non-convex nature of the posterior probability. The posterior probability is proportional to the product of likelihood and hyper prior. For the kernel length scales, a gamma distribution as well as a uniform distribution with smooth edges is considered as the hyper prior. All GPR models are generated using the GPML toolbox \cite{Rasmussen.2010}.

\subsection{Acquisition function} \label{sec:AccMdls}

Maximizing the acquisition function determines the location of the next query point (cf. Algo. 1: Step 5) by balancing exploitation and exploration. 
In this contribution, three different  
acquisition functions are compared. 

Firstly, the classical Expected Improvement (EI) \cite{Jones.1998} is used. 
It maximizes the expectation of the amount of improvement w.r.t. the best point found so far.   
Secondly, the Upper Confidence Bound (UCB) \cite{Auer.2002} is used. The idea is to optimistically search for the next sample point by (in the case of minimization) subtracting a scaled standard deviation from the expected mean: $\alpha(\boldsymbol{\theta}) = \mu_J \left( \boldsymbol{\theta}\right) - \beta \sigma_J\left(\boldsymbol{\theta}\right)$.  
Here, we use a constant value of $\beta = 3$. The value of $\beta$ was not tuned to the test cases at hand. Thirdly, Max-Value Entropy Search (MES) \cite{Wang.06.03.2017} is used. MES is considered an information theoretic acquisition function. It determines the next query point, by maximizing the information gain about the distribution of the unknown objective function value of the optimum.

In all cases, the acquisition function is maximized by using a combination of random search and gradient based optimization because it can be highly multi modal. 

\subsection{LCC using Virtual Data Points (VDP)}  \label{sec:crashConstraints}

As described in Sec. \ref{sec:Problem}, tuning problems in control engineering can be subject to crash constraints. Infeasible parameters lead to the objective function value being not available or extremely large. 
Setting $J$ to some arbitrary fixed value 
may be not a valid option because this may result in a discontinuous objective function which contradicts the assumptions encoded in the kernel (cf. e.g. \cite{Marco.2021, Dorschel.2021}, Fig. \ref{fig:wocrash}). Additionally, due to the discontinuity at $\theta = 0.85$, the GP hyperparameter optimization yields a small length scale, which deteriorates GP performance globally.
Here, we propose a simple heuristic approach, which 
only modifies the training data for the GP and therefore can easily be included in other BO extensions. For example, the procedure was introduced superficially for constrained optimization in our recent conference paper \cite{Stenger.2022}.

\begin{algorithm}[h] 
	1: Extract all crashed evaluations: \hbox{$\bar{\mathcal{D}}_k  := (\bar{\Theta}_k, \bar{\mathcal{J}}_{k})$},\\
	\hspace*{3.2mm} and all successful evaluations: \hbox{$\breve{\mathcal{D}}_k  := (\breve{\Theta}_k, \breve{\mathcal{J}}_{k})$.}   \\ [3pt]
	2: Fit GP Model with all successful evaluations $\breve{\mathcal{D}}_k$. \\ [3pt]
	3: \textbf{for each} crashed query $\bar{\boldsymbol{\theta}}_i \in \bar{\Theta}_k$  \\ [3pt]
	4: \quad Make probabilistic prediction using $\breve{\mathcal{D}}_k$: 
	\begin{equation*}
		\tilde{J}(\bar{\boldsymbol{\theta}} \mid \breve{\mathcal{D}}_k) \sim \mathcal{N} \left(\breve{\mu}_J (\bar{\boldsymbol{\theta}_i}), \breve{\sigma}_J^2 (\bar{\boldsymbol{\theta}_i})  \right)
	\end{equation*}
	5: \quad Calculate virtual data point using a pessimistic GP \\
	\quad \hspace*{6.5mm}  prediction:   $\hat{J}_i = \breve{\mu}_J (\bar{\boldsymbol{\theta}_i}) + \gamma \breve{\sigma}_J (\bar{\boldsymbol{\theta}_i})$  \\ [3pt]
	6: \quad Bound the pessimistic prediction to the range of \\  
	\quad \hspace*{6.5mm} past successful evaluations: \\
	\quad \hspace*{6.5mm} $\hat{J}_i = \mathrm{min}\{\mathrm{max}\{\hat{J}_i(\breve{\boldsymbol{\theta}}), J_\mathrm{min}\} + \gamma \breve{\sigma}_J\ , J_\mathrm{max}\}$  \\ 
	\caption{Calculation of Virtual Data Points (Step 3 of \hbox{Algo. 1})}
	\label{Algo:CrashConstraint}
\end{algorithm}

The approach is summarized in Algo. \ref{Algo:CrashConstraint}. The idea is to add virtual data points at the location of the crashed objective function queries, where $l(\boldsymbol{\theta}) = 0$. The goal is to choose the artificial objective function values such that, the optimization is pushed away from  
the crashed query without contradicting the smoothness assumptions encoded in the GP. 

First, the GP is fitted using only successful queries (Steps 1, 2).  
Afterwards, probabilistic predictions are made 
for each of the failed queries (Step 4). These probabilistic predictions are used to calculate a pessimistic realization (Step 5). The additional hyperparameter $\gamma$ is set to $\gamma = 3$. It was not tuned to the test cases at hand. 
As the last step, the virtual objective function value $\hat{J}_i$ is constrained such that  $  J_\mathrm{min} + \gamma \breve{\sigma}_J \le \hat{J}_i \le J_\mathrm{max}$ holds.  
The minimum and maximum successfully obtained objective function values are denoted as $J_\mathrm{min}$ and $J_\mathrm{max}$. The lower bound ensures that areas around crashed locations are not considered best and therefore, helps to prevent repetitive sampling in the infeasible region. Constraining to the observed maximum is useful in case of large predictive uncertainties. This procedure is repeated for all failed evaluations at each BO iteration (Step 3). 
\begin{figure}[th]
	\centering
	\includegraphics[width = 0.45 \textwidth]{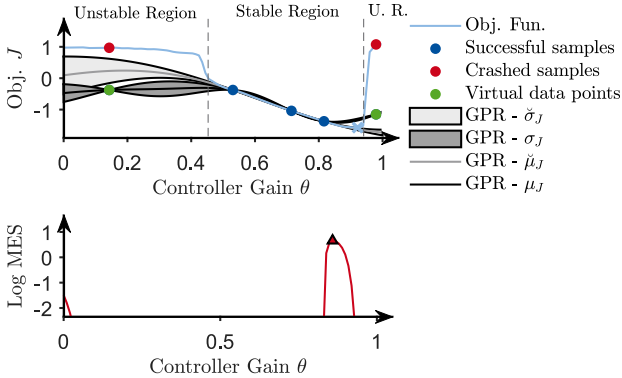}
	\caption{BO example with crash constraint handling using VDP. \textbf{Top}: Objective function, evaluations, and GP models. \textbf{Bottom}: Acquisition function.}
	\label{fig:wcrash}
\end{figure}

Fig. \ref{fig:wcrash} illustrates the approach. In contrast to Fig. \ref{fig:wocrash}, the length scale is increased due to the less unlikely virtual data points - the green points are closer to the original prediction than the red points. As a result, the predictive uncertainty is greatly reduced. Additionally, the prediction for $\theta > 0.8$ does increase more slowly. This results in the maximum of the acquisition function marked by the red triangle to be located closer to the minimum of the ground truth than in Fig. \ref{fig:wocrash}.

\section{Controller Tuning Benchmark Problems} \label{sec:Benchmark}

\subsection{Test Cases} \label{sec:TestCases}

In order to meaningfully compare the sample efficiency of
the different optimizers, ten different tuning problems from the field of control engineering are presented. Their characteristics are summarized in Table \ref{tab:BenchTabel}. The algorithms to be tuned range from LQR, via MPC to an unscented Kalman filter (UKF). Low level objective functions such as  integral of time-multiplied absolute value of error (ITAE), e.g. test case 2 and 8, as well as higher level objectives such as passenger comfort (test case 6 and 10) are addressed. The problem dimensionality $d$ ranges from two to five.
Test cases 1, 2, 3, 7, and 8 can be considered toy problems, whereas the remaining test cases origin from applied control engineering research. For the toy problems, the objective function evaluation time is less than one second. The other problems have objective function evaluation times of several seconds. In all but three test cases, simulation crashes can occur.
Below, the test cases are described in more detail. It is also explained for each test case how the returned objective function value is calculated in case of crashed simulations, if VDP is not used.

\begin{table*}[th] 
	\caption{Summary of the ten test cases with approximate relative size of the global optimum $s_\mathrm{opt}$, frequency of simulation crashes $p_{\mathrm{crash}}$ and objective function evaluation time $T_\mathrm{sim}$ (cf. Sec. \ref{sec:ObjFunAna}) }
	\centering
	\begin{tabular}{p{1cm}| >{\raggedleft}p{2cm} >{\raggedleft}p{2cm} >{\raggedleft}p{1cm} >{\raggedleft}p{2.3cm} >{\raggedleft}p{2cm} | >{\raggedleft}p{1cm} >{\raggedleft}p{1cm} >{\raggedleft}p{1cm}  }
		\toprule
		Test case No. & Application & Algorithm(s)   &  Dimension $d$          &  Objective Function    &  Reasons for simulation crash & $s_{\mathrm{opt}}$  & $p_{\mathrm{crash}}$  & $T_\mathrm{sim}$ 
		\tabularnewline
		\midrule
		1 & Balancing robot \cite{Framing.2020}                       & LQR              &  $2$            & Tracking accuracy (RMSE)  & Robot falls over & $49\%$ &  $45\%$	 & $0.4 \, s$  
		\tabularnewline
		\rule{0pt}{3ex}%
		2 & Inverted pendulum   &  State feedback controller              & 2            & ITAE & Swing down & $92 \%$ & $20 \%$  &  $0.7 \, s$  	   
		\tabularnewline %
		\rule{0pt}{3ex}%
		3 & Coupled tank system \cite{Scheurenberg.2022}   &  PI-controller           & 2            & Tracking \& overflow prevention  & - & $54 \%$ &  $0 \%$ & $< 0.1 \, s$  	   
		\tabularnewline%
		\rule{0pt}{3ex}%
		4 & Milling tool velocity \cite{Stenger.2020}   &  MPC \& EKF              & 2            & MAE \& Overshoot barrier  & - & $8\%$ &  $0 \%$ & $26.3 \, s$   	   
		\tabularnewline %
		\rule{0pt}{3ex}%
		5 & Underwater vehicle \cite{Stenger.2022}   &  Path planner \& Guidance              & 3            & Energy consumption  & AUV fails to reach goal & $7 \%$ &  $52 \%$  & $40.7 \, s$  	   
		\tabularnewline %
		\rule{0pt}{3ex}%
		6 & Active vehicle damping   &  Skyhook + ADD controller   & 3 & Riding comfort \& Wheel forces   & - & $58 \%$ &  $0 \%$ & $5.3 \, s$   	   
		\tabularnewline %
		\rule{0pt}{3ex}%
		7 &  Balancing robot \cite{Framing.2020}                       & LQR              &  $4$            & Tracking Ac- curacy (RMSE)  & Robot falls over & $7 \%$ &  $45  \%$ & $0.4 \, s$ 
		\tabularnewline
		\rule{0pt}{3ex}%
		8 & Inverted pendulum   &  State feedback controller              & $4$            & ITAE & Swing down & $70\%$ & $35\%$
	    & $0.7 \, s$      	   
		\tabularnewline %
		\rule{0pt}{3ex}%
		9 & Underwater vehicle \cite{Stenger.2022}   &  UKF              & $4$            & $90-$Percentile NED error  & Filter diverges & $88\%$ & $27\%$ & $20.3 \, s$  	   
		\tabularnewline %
		\rule{0pt}{3ex}%
		10 & Autonomous driving \cite{Ritschel.2019, GharibAli.2021}   &  Path following MPC  & 5            & Comfort \& Safety  & Vehicle deviates from road & $95 \%$ & $7 \%$ & $14.3 \, s$   \tabularnewline	   
		\bottomrule
	\end{tabular}\label{tab:BenchTabel}
\end{table*}

\paragraph{Test Case 1}
 
The plant is a balancing robot with two wheels called EDU-BAL. The task is to tune the parameters of an LQR controller. Details of robot model and controller can be found in \cite{Framing.2020}. For each objective function evaluation with duration $T$, two step changes are applied to the lateral position reference $x_\mathrm{ref}$ of the robot wheels. Objective function $J_1$ is the RMSE of the position $x(t)$:

\begin{equation}
	J_1 =  \sqrt{\frac{1}{T}\int_0^{T} \left(x_{\mathrm{ref}}(t) - x(t) \right)^2 dt}.
\end{equation}

Optimization variables are $Q_1$ penalizing position error and $Q_{3,4}$ penalizing the derivatives of position and body angle: \hbox{$\boldsymbol{\theta} = [Q_1 \ Q_{3,4}]$}. The remaining weights 
are kept constant. Parametric model plant mismatch and a sensor bias are introduced. As a result, for specific LQR parameters the closed loop becomes unstable. A simulation crash is defined as the body angle exceeding $\pm 90^{\circ}$. If the VDP method is not used, a fixed objective function is assigned in the case of a crash.

\paragraph{Test Case 2}

The feedback gains
of a state feedback controller are optimized for a non-linear cart-pole system.  
Objective is to track the reference trajectory $x_{\mathrm{ref}}$ for the position $x$ of the cart consisting of two differently sized reference steps (denoted as 1 and 2) and at the same time minimizing the pole angle $\varphi$. The objective function 

\begin{equation}
	J_2 = 0.45(\mathrm{ITAE}_{\varphi,1} + \mathrm{ITAE}_{\varphi,2}) + 0.05(\mathrm{ITAE}_{x,1} + \mathrm{ITAE}_{x,2}),
\end{equation}

is based on the ITAE criterion: $	\mathrm{ITAE} = \int_0^T t \, \|e\| \,\mathrm{d}t$.
The feedback gains corresponding to cart position and velocity are kept constant. Optimization variables are the feedback gains corresponding to angle and angular velocity: \hbox{$\boldsymbol{\theta} = [k_{\varphi} \ k_{\dot{\varphi}}]$}. A simulation crash is defined as $\varphi$ exceeding $\pm 90^{\circ}$. In this case the objective function value is set to the open-loop performance.      

\paragraph{Test Case 3}

The classical three-tank system is considered. The non-linear equations for the simulation of the three-tank-system are described in \cite{Scheurenberg.2022}. Objective is to track a reference step $V_{3,ref}$ for the water level of the third tank $V_3$, while avoiding critical water levels in the first $V_1$ and second  $V_2$ tank:

\begin{equation}
	\begin{aligned}
			J_3 = &\frac{1}{T}\sqrt{\int_{0}^T \left( V_{3,\mathrm{ref}}(t) - V_3(t)\right)^2 dt} +  \\
			 &10 \, \max \, \{0, V_2(t) - 5.5,V_1(t) - 8 \}.
	\end{aligned}
\end{equation}

 The tank level in the third tank is controlled by actuating a pump which determines the volume inflow towards the first tank using a PI Controller with $\boldsymbol{\theta} = [k_p, k_i]$. Valves between the tanks are not actuated.

\paragraph{Test Case 4}

The control of the velocity $v$ of a milling tool is considered. Control objective is to minimize the average absolute tracking error $|e|_{v,n}$ for multiple steps in the reference velocity of the tool. Critical overshoot of the tool velocity in positive direction $\Delta v_{\mathrm{max},n}$ should be avoided: 

\begin{equation}
	J_4 = \frac{1}{5} \sum_{n=1}^5 |e|_{v,n} +\frac{1}{1+ 50 \, \mathrm{exp}(-\Delta v_{\mathrm{max},n} + 0.3)} \, .
\end{equation}  

For each parametrization, the control performance is averaged over five different model plant missmatches and sensor noise seeds $n$. 
The second term constitutes a barrier function for the overshoot constraint.
The system is controlled using a linear MPC and states are estimated using an EKF. Tuning parameters are the ratios $\lambda$ between the scalar $Q$ and $R$ of the MPC and EKF: \hbox{$\boldsymbol{ \theta} = [\lambda_{\mathrm{MPC}}, \lambda_{\mathrm{EKF}}]$}. For more information on plant and controller, the reader is referred to \cite{Stenger.2020}.

\paragraph{Test Case 5} The path planning and guidance algorithms of a miniature autonomous underwater vehicle (AUV) is optimized. Control objective is to visit different way-points with as little energy consumption as possible. The energy consumption is calculated for three random seeds $n$ of different currents, model-plant mismatches, way points, and sensor noise. In order to calculate the objective function, the average energy consumption is taken:

\begin{equation}
	J_5 = \frac{1}{3} \sum_{n=1}^3 \int_0^T P_n (t) \mathrm{d} t \, . 
\end{equation}

Optimized parameters are the minimum turning radius of the path planner $r_\mathrm{plan}$, the reference velocity $v_\mathrm{plan}$ and the look-ahead distance $\Delta_\mathrm{LOS}$ of the guidance module: \hbox{$\boldsymbol{ \theta} = [r_\mathrm{plan}, v_\mathrm{plan}, \Delta_\mathrm{LOS}]$}. Details 
can be found in \cite{Stenger.2022}. A simulation crash is defined, as the AUV not being able to reach the next way point. In this case a fixed objective function value is assigned, if VDP is not used.

\paragraph{Test Case 6}

The controller for an active vehicle suspension for a linear full vehicle model is optimized. As a controller a variant of the so called Skyhook + ADD \cite{Savaresi.2006} is used. To narrow the sim-to-real gap, measured road excitation is used. The RMSE of the tire forces $E_\mathrm{tire}$ representing safety and the acceleration of a passenger filtered according to ISO2631 $E_\mathrm{acc}$ representing comfort is minimized. 
As a reference, the performance characteristics of a passive damper $E_\mathrm{tire,ref}$ and $E_\mathrm{acc,ref}$ are used. The objective function is chosen as the average ratio of the performance criteria, where worse than passive performance is penalized quadratically:  
 
\begin{equation}
	J_6 = 1/2 \Delta_\mathrm{tire} + 1/2 \Delta_\mathrm{acc}, 
\end{equation}

with

\begin{equation}
	\Delta_* = 
	\begin{cases}
		\frac{E_*}{E_{*,\mathrm{ref}}}, & \text{if } E_* < E_{*,\mathrm{ref}} \\
		2\frac{E_*}{E_{*,\mathrm{ref}}} + \frac{E_*}{E_{*,\mathrm{ref}}} ^2 ,              & \text{otherwise} \ .
	\end{cases} 
\end{equation}

The controller has three parameters $\boldsymbol{\theta} = [k_1 \ k_2 \ k_3]$. A similar problem setting was solved experimentally in \cite{GianlucaSavaia.2021}.

\begin{figure*}[th]
	\centering
	\includegraphics[width=\textwidth]{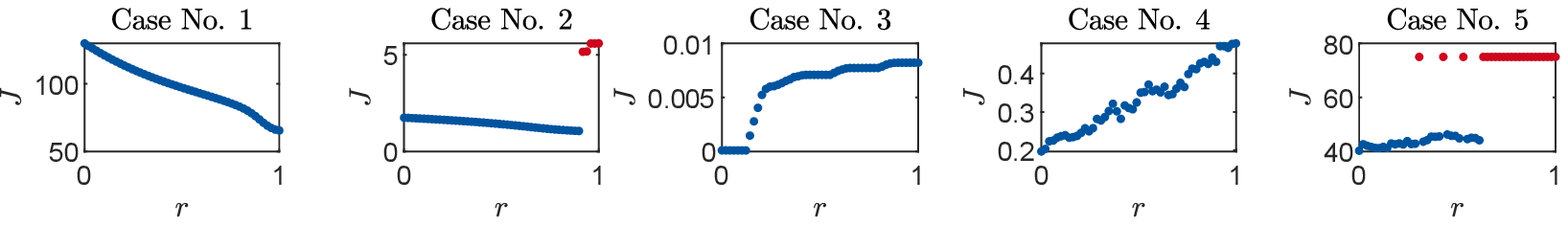}
	
	\hspace{0.5cm}
	
	\includegraphics[width=\textwidth]{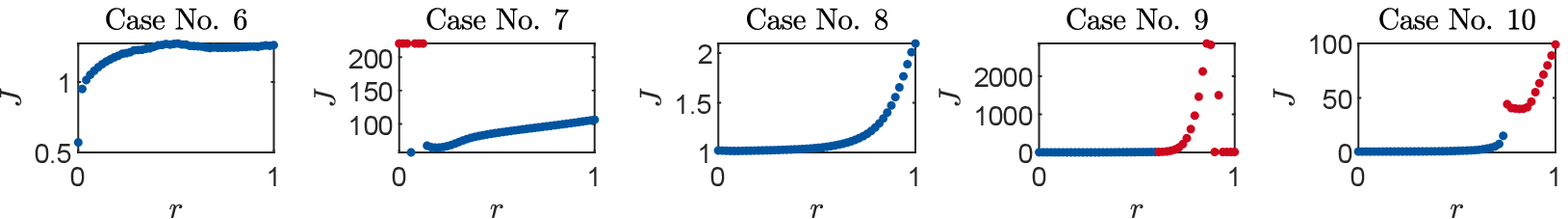}
	\caption{Visualization of the objective function landscape. One of the 1-d subspaces are shown for each test case. Crashed simulations are marked in red.}
	\label{fig:objFunAna}
\end{figure*}

\paragraph{Test Case 7}

Test case 7 is identical to test case 1 (balancing robot) with the exception that all entries of $\mathbf{Q}$ are optimized: \hbox{$\boldsymbol{\theta} = [Q_1 \ Q_2 \ Q_3 \ Q_4]$}.

\paragraph{Test Case 8}

Test case 8 is identical to test case 2 (inverted pendulum) with the exception that all gains of the state feedback controller are optimized: \hbox{$\boldsymbol{\theta} = [k_x, k_{\dot{x}},k_{\varphi} \ k_{\dot{\varphi}}]$}.

\paragraph{Test Case 9}

The optimization of the navigation filter for the same AUV as in test case 5 is considered. Four parameters of an UKF are optimized (cf. Sec. IV-A in \cite{Stenger.2022}): \hbox{$\boldsymbol{\theta} = [\alpha_1 \ \alpha_2 \ \alpha_3 \ \alpha_4]$}. Objective is the $90$-percentile of the north-east-down (NED) position estimation error on a $650 \ s$ long fixed vehicle trajectory:

\begin{equation}
	J_9 = Q_{90} (e_{\mathrm{NED}}) \ .
\end{equation}   

A simulation crash is defined, as the NED-error exceeding $10\ \mathrm{m}$ or ill-conditioned covariances occurring. In the first case, the objective function is returned as is. In the second case, the objective function value is fixed to $10 \ \mathrm{m}$.     

\paragraph{Test Case 10}
An MPC for vehicle path following in autonomous driving is optimized. Similarly to test case 6, competing objectives are considered. In this case, they are the velocity RMSE $E_\mathrm{v}$, lateral distance RMSE $E_\mathrm{lat}$, and the RMSE of the acceleration $E_\mathrm{acc}$. The objective function is formulated as:

\begin{equation}
	J_{10} = 1/3 \Delta_\mathrm{v} + 1/3 \Delta_\mathrm{lat} + 1/3 \Delta_\mathrm{acc} , 
\end{equation}

with

\begin{equation}
	\Delta_* = 
	\begin{cases}
		\frac{E_*}{E_{*,\mathrm{ref}}}, & \text{if } E_* < E_{*,\mathrm{ref}} \\
		\frac{E_*}{E_{*,\mathrm{ref}}} + \frac{2E_*}{E_{*,\mathrm{ref}}} ^2 ,              & \text{otherwise} \ .
	\end{cases} 
\end{equation}

Three elements of the MPC's $\mathbf{Q}$ and two elements of the MPC's $\mathbf{R}$ are tuned:  \hbox{$\boldsymbol{\theta} = [Q_1 \ Q_2 \ Q_3 \ R_1 \ R_2]$}. A simulation crash is defined as the vehicle not reaching the goal state or the lateral error exceeding a  
threshold. In case of a crash the data obtained until the crash is used to calculate the objective function value.   
The considered MPC is introduced in \cite{Ritschel.2019}. In \cite{GharibAli.2021}, a multi-objective optimization approach 
is presented.

\subsection{Objective Function Analysis} \label{sec:ObjFunAna}

\paragraph{Method} In order to characterize the objective function landscape of each of the test cases, we create random one dimensional subspaces as follows:

\begin{equation}
	\hat{\boldsymbol{\theta}} (r) =  \boldsymbol{\theta}^* + (r + r_0) \mathbf{A} \quad \mathbf{A} \in \mathbb{R}^{d \times 1}.
\end{equation}

The vector $\mathbf{A}$ defines a randomly chosen direction in the original design parameter space. The scaling of $\mathbf{A}$ and the value $r_0$ are chosen such that $\hat{\boldsymbol{\theta}} (r=0)$ and $\hat{\boldsymbol{\theta}} (r=1)$ are located on the border of the original domain defined by the box constraints in Eq. \ref{eq:BO_OptProb}. All subspaces go through the global optimum of the respective test case $\boldsymbol{\theta}^*$\footnote{The true optimal value is unknown. Here we chose the best evaluation recorded during all optimizations.}. Ten random subspaces are created for each test case. From each of them 51 parametrizations are evaluated in an equidistant fashion with $r = 0, \ 0.02\, \dots$ . The following quantities are evaluated: 

\begin{itemize}
	\item Percentage of crashes $p_{\mathrm{crash}}$: Fraction of crashed evaluations, where $l(\hat{\boldsymbol{\theta}} (r)) = 0$. 
	\item The average time for one objective function \hbox{evaluation $T_{\mathrm{sim}}$}.
	\item Relative size of the global Optimum $s_{\mathrm{opt}}$:
	The percentage of evaluations, which do not have a local maximum located on a straight line between them and the global minimum $\boldsymbol{\theta}^*$. Because the objective function is unknown, we cannot directly count the number of local minima. Therefore, this measure is chosen 
	to give an intuition about the multimodality.   
\end{itemize}     
\vspace{\baselineskip}    

\paragraph{Results} Fig. \ref{fig:objFunAna} shows one of the respective subspaces for each test case. Table \ref{tab:BenchTabel}, right, lists the corresponding quantitative results. The controller tuning objective functions can be characterized as follows:

(1): Substantial fractions of the parameter space are subject to crashes. 
	At the borders between the crashing and successful portions of the parameter space, discontinuities or large gradients can be observed.	
	
(2): The objective function can be corrupted by deterministic noise. For example test cases 4 and 5 exhibit a trend towards better behavior if $r$ decreases. This trend is superposed by fluctuations on small parameter scales resulting in the relative size of the global optimum $s_{\mathrm{opt}}$ becoming small. As an explanation, substantial deterministic artificial sensor noise is included in the simulations of test cases 4 and 5. 	

(3):  Apart from the deterministic noise and test case 6, the subspaces are mostly dominated by one global optimum. For a discussion of the narrow global optimum of test case 7, see Sec. \ref{sec:TimeDomainAna}.

\section{Benchmark Results} \label{sec:res}

\subsection{Evaluated Algorithms} \label{sec:BenchAlgos}

\paragraph{BO Nomenclature}
In order to distinguish between the different BO variants presented in Sec. \ref{sec:BO}, we use the following notation: \\

\{MES,UCB,EI\}-\{SE, MA\}\{Q,G\}-\{F,V\}. \\

As the default BO variant, we use the squared exponential (SE) kernel with Max-Value Entropy Search (MES) as the acquisition function, and no specific treatment for crash constraints (F): MES-SE-F. Instead of MES, the Upper Confidence Bound (UCB) and Expected Improvement (EI) acquisition functions are employed. In addition, we use a quadratic mean function (Q) and also consider a gamma hyperprior (G). Either the default treatment (F) as explained in Sec. \ref{sec:TestCases} for each test case, i.e. assigning a fixed value or using the data obtained until the crash, or the VDP method (V) is chosen to handle crash constraints. In total, nine combinations are evaluated \hbox{(cf. Table \ref{tab:allResults})}. 

\paragraph{Benchmark Optimizers} Numerous different algorithms 
have been used for automated tuning in control engineering (cf. Sec. \ref{sec:relwork}). Here we choose the well-known metaheuristics particle swarm optimization (PSO),  
genetic algorithm (GA), 
and covariance matrix adaptive evolutionary search (CMAES) \cite{Hansen.2001}. In addition to that, pattern search (PS) 
is used. BADS \cite{Acerbi.2017} combines PS and BO and therefore is expected to work well in the LCC setting, because it has a fallback strategy in case the GP model is corrupted by crashed evaluations. Because of that fallback strategy and the promising performance of PS, BADS is included in the benchmark.  
Additional baselines are random search (Rand) and full-factorial design space exploration (Grid). For Grid, each dimension is discretized into $\lceil \mathrm{log}_d (25 \, d) \rceil$ equally spaced levels. Afterwards all parameter combinations are evaluated. As a result, the full factorial search has a larger budget than the budget of the other algorithms ($25 \, d$) with the exception of $d = 5$, where $\mathrm{log}_5 (125) = 3$. 
In order to verify our implementation of BO, Matlabs own BO implementation (BayesOpt) is chosen as an additionally benchmark. Lastly, Fmincon,  one of MATLABs standard locally searching optimizers, is used.
For PSO, GA, PS, BayesOpt and Fmincon the build-in implementations of MATLAB version 2020b are used. The implementations of CMAES and BADS are taken from public repositories\footnote{CMAES:  \url{https://de.mathworks.com/matlabcentral/fileexchange/52898-cma-es-in-matlab}}\footnote{BADS: \url{https://github.com/lacerbi/bads}}. 
The default hyperparameters of the respective software packages are used, without fine-tuning to the test cases at hand.

\subsection{Performance Metrics} \label{sec:perfMetrics}

Each optimizer is run multiple times for each test case in order to evaluate whether differences in performance are statistically significant. For the toy problems (1, 2, 3, 7 and 8), each algorithm is run 50 times with different seeds for the initial sampling. For the other more expensive-to-evaluate real world problems, each algorithm is run ten times. Note that for one seed the initial sampling is identical for all compared optimizers. 
The following performance metrics are used:

\paragraph{Average Rank} Starting point for the calculation of the average rank is the rank 
of the various optimizers for each seed and each test case. The rank is calculated by sorting the optimizers w.r.t. the best found objective function value after a given budget (e.g. $5 \, d$). The best performing optimizer has rank one. The second best has rank two etc. Since the rank of the performance of the algorithms depend on the initial sampling seed, the rank is averaged over the different seeds. 
Afterwards, the rank is averaged over the different test cases ensuring that each test case contributes equally. 

\paragraph{Scaled Regret} The rank does not take the magnitude of the difference in performance into account. Therefore, the simple regret $r_n(s,k,o) = J_{n,\mathrm{min}}(s,k,o) - J_n (\boldsymbol{\theta}^*)$
for test case $n$, relative budget $k$ (e.g. $5 \, d$), optimizer $o$ and seed $s$ is used. It is calculated by subtracting the global optimum $J_n (\boldsymbol{\theta}^*)$\footnote{The true optimal value is unknown. Here we chose the best evaluation recorded over all function evaluations.}  
from the best solution found so far $J_{n,\mathrm{min}}(s,k,o)$.  
In order to make the regret comparable across different test cases we scale it with respect to the median regret of random search at evaluation $25 \, d$: 

\begin{equation}
	r_{\mathrm{s},n}(s,k,o) = \frac{r_n(s,k,o)}{\mathrm{median}(r_n(s,25 \, d,Rand))} 
\end{equation}

This ensures that the result is invariant to objective function scaling. The median $\bar{r}_{\mathrm{s},n}(s,k,o)$ and the 
$80 \%$ quantile of the scaled regret 
$r_{\mathrm{s},80 \%, n}(s,k,o)$ are calculated for each optimizer, test case and budget. Both are afterwards averaged over the different test cases.   

\paragraph{Statistical Significance}\label{sec:perfMetricsStat}

In order to test for statistical significance of differences in optimizer performance (scaled regret), a hypothesis test is used for each test case and a budget of $25 \, d$. It cannot be assumed that the scaled regret follows a Gaussian distribution. Therefore, the non-parametric one sided Wilkoxon rank sum test is used with a significance level of $5 \ \%$.

\newcommand{\myTE}[4] {#1. \hspace{0.1cm}   #2}
\newcommand{\myTEg}[4] {\textcolor{gray}{#1. \hspace{0.1cm}   #2}}
\newcommand{\myTEb}[4] {\textbf{#1. \hspace{0.1cm}   #2}}
\newcommand{\myTEbg}[4] {\textcolor{gray}{\textbf{#1. \hspace{0.1cm}   #2}}}
\newcommand{\myTEbb}[4] {\textbf{\underline{#1. \hspace{0.1cm}   #2}}}
\newcommand{\myTEl}[4] {#1. \hspace{0.1cm}   #2}

\begin{table*}[th] 
	\caption{Benchmark results after $25 \, d$ evaluations. The median scaled regret $\bar{r}_{\mathrm{s},n}(s,25 \, d,o)$,  
	(cf. Sec. \ref{sec:perfMetrics}) is shown. Results, which are not statistically significantly worse than the best optimizer (underlined) are marked in bold. Median scaled regrets of more than one (i.e. worse than random search) are marked in light gray.}
	\centering
	\begin{tabular}{p{1.5cm}| r r r r r r r r r r}
		\toprule
		& \multicolumn{10}{c}{Test case No.}
		\tabularnewline
		Algorithm & 1 ($d = 2$)    &  2 ($d = 2$)          &  3 ($d = 2$)    & 4 ($d = 2$)  & 5 ($d = 3$)  & 6 ($d = 3$) &7 ($d = 4$) &8 ($d = 4$) &9 ($d = 4$) &$10$ ($d = 5$)
		\tabularnewline
		\midrule
		Grid & \textcolor{gray}{1.6} & 0.19 & 0.25 & 0.88 & 0.95 & \textcolor{gray}{1.3} & \textcolor{gray}{3.35} & 0.56 & \textcolor{gray}{1.14} & 0.42 \tabularnewline		 		
		Rand & \myTE{17}{1.00}{0.07}{3.02} & \myTE{14}{1.00}{0.37}{1.63} & \myTE{16}{1.00}{0.25}{9.45} & \myTE{12}{1.00}{0.24}{1.39} & \myTE{14}{1.00}{0.71}{1.18} & \myTE{15}{1.00}{0.83}{2.21} & \myTE{16}{1.00}{0.86}{1.38} & \myTE{13}{1.00}{0.40}{1.44} & \myTEb{3}{1.00}{0.54}{1.90} & \myTE{15}{1.00}{0.45}{1.97} \tabularnewline[0.05cm] 
		\midrule 	
		MES-SE-F & \myTE{13}{0.25}{0.02}{1.88} & \myTE{11}{0.72}{0.52}{1.42} & \myTE{10}{0.27}{0.09}{0.62} & \myTE{5}{0.40}{0.11}{0.71} & \myTEbb{1}{0.28}{0.08}{0.56} & \myTE{12}{0.12}{0.01}{0.13} & \myTE{9}{0.93}{0.87}{1.07} & \myTE{10}{0.42}{0.22}{0.71} & \myTEg{12}{1.72}{1.52}{2.69} & \myTE{14}{0.72}{0.13}{1.25} 	\tabularnewline[0.05cm] 	
		UCB-SE-F & \myTE{14}{0.30}{0.06}{2.03} & \myTE{10}{0.68}{0.48}{1.30} & \myTE{3}{0.16}{0.06}{0.38} & \myTE{10}{0.63}{0.22}{0.99} & \myTEb{4}{0.40}{0.12}{0.47} & \myTE{9}{0.06}{0.02}{0.17} & \myTE{13}{0.97}{0.87}{1.09} & \myTE{5}{0.34}{0.21}{0.64} & \myTEg{16}{3.95}{0.78}{6.01} & \myTEb{10}{0.51}{0.12}{1.16}	\tabularnewline[0.05cm] 
		EI-SE-F & \myTE{9}{0.10}{0.03}{0.36} & \myTE{9}{0.51}{0.32}{0.67} & \myTE{5}{0.20}{0.11}{0.44} & \myTE{7}{0.51}{0.22}{0.93} & \myTEb{5}{0.42}{0.22}{0.72} & \myTE{8}{0.05}{0.02}{0.10} & \myTE{7}{0.92}{0.86}{0.99} & \myTE{7}{0.37}{0.20}{0.51} & \myTEg{11}{1.71}{1.06}{2.64} & \myTEb{3}{0.17}{0.14}{0.87}
		\tabularnewline[0.05cm] 
		
		MES-MA-F & \myTE{11}{0.22}{0.02}{1.98} & \myTE{12}{0.82}{0.50}{1.66} & \myTE{12}{0.44}{0.14}{0.89} & \myTE{6}{0.40}{0.22}{1.30} & \myTEb{3}{0.39}{0.21}{0.67} & \myTE{6}{0.05}{0.02}{0.12} & \myTE{6}{0.90}{0.86}{0.99} & \myTE{9}{0.42}{0.21}{0.75} & \myTEg{14}{1.81}{1.44}{2.76} & \myTE{12}{0.68}{0.37}{1.28}
		\tabularnewline[0.05cm] 
		
		MES-SE-V & \myTE{6}{0.06}{0.01}{0.17} & \myTE{5}{0.13}{0.02}{0.68} & \myTE{9}{0.24}{0.11}{0.66} & \myTEg{13}{1.02}{0.31}{1.14} & \myTEb{6}{0.43}{0.14}{0.57} & \myTE{4}{0.04}{0.02}{0.14} & \myTEb{2}{0.87}{0.59}{0.98} & \myTE{8}{0.38}{0.22}{0.91} & \myTEbg{5}{1.08}{0.42}{1.44} & \myTEbb{1}{0.12}{0.11}{0.25}
		\tabularnewline[0.05cm] 
		
		EI-SE-V & \myTE{10}{0.14}{0.02}{0.49} & \myTE{4}{0.08}{0.02}{0.44} & \myTE{7}{0.20}{0.09}{0.48} & \myTE{9}{0.62}{0.12}{1.38} & \myTE{8}{0.47}{0.26}{0.77} & \myTE{11}{0.09}{0.06}{0.16} & \myTEb{3}{0.88}{0.85}{0.92} & \myTE{4}{0.30}{0.14}{0.59} & \myTEbg{4}{1.03}{0.67}{1.27} & \myTE{11}{0.63}{0.16}{0.96}
		\tabularnewline[0.05cm] 
		
		MES-MA-V & \myTEb{2}{0.00}{0.00}{0.06} & \myTE{2}{0.01}{0.00}{0.40} & \myTE{11}{0.33}{0.18}{0.75} & \myTE{4}{0.39}{0.20}{0.95} & \myTEb{9}{0.53}{0.20}{0.81} & \myTE{5}{0.04}{0.01}{0.10} & \myTEbb{1}{0.86}{0.85}{0.94} & \myTE{6}{0.36}{0.23}{0.70} & \myTEg{8}{1.42}{1.03}{1.79} & \myTEb{4}{0.19}{0.12}{0.82}
		\tabularnewline[0.05cm] 
		
		MES-SEQ-V & \myTE{12}{0.25}{0.03}{1.30} & \myTE{7}{0.29}{0.06}{1.88} & \myTE{4}{0.17}{0.07}{0.56} & \myTE{11}{0.93}{0.23}{3.64} & \myTEb{2}{0.37}{0.10}{0.65} & \myTE{3}{0.04}{0.03}{0.06} & \myTE{8}{0.92}{0.86}{1.19} & \myTEg{16}{1.70}{0.76}{3.08} & \myTEg{17}{5.20}{0.72}{9.61} & \myTEb{7}{0.29}{0.12}{0.58}
		\tabularnewline[0.05cm] 
		
		MES-SEG-V & \myTE{7}{0.06}{0.01}{0.21} & \myTE{6}{0.18}{0.02}{0.59} & \myTE{8}{0.22}{0.09}{0.43} & \myTE{8}{0.57}{0.21}{1.13} & \myTE{10}{0.57}{0.42}{0.87} & \myTE{10}{0.06}{0.03}{0.11} & \myTE{15}{0.98}{0.89}{1.10} & \myTE{11}{0.81}{0.37}{1.59} & \myTEbb{1}{0.80}{0.22}{1.60} & \myTEg{16}{1.22}{1.13}{1.42}
		\tabularnewline[0.05cm] 
		\midrule 
		
		GA & \myTE{4}{0.02}{0.00}{3.75} & \myTEg{16}{4.27}{1.30}{8.40} & \myTE{13}{0.56}{0.09}{116} & \myTEg{16}{3.34}{1.02}{0.94} & \myTEg{15}{1.07}{0.95}{1.63} & \myTEg{16}{4.56}{2.27}{6.56} & \myTE{12}{0.97}{0.86}{1.00} & \myTEg{15}{1.28}{0.47}{3.73} & \myTEg{10}{1.71}{1.10}{5.59} & \myTE{13}{0.71}{0.57}{1.04}
		\tabularnewline[0.05cm] 
		
		CMAES & \myTE{15}{0.38}{0.10}{1.92} & \myTE{13}{0.85}{0.33}{2.67} & \myTE{15}{0.87}{0.21}{0.94} & \myTEg{14}{1.05}{0.83}{1.34} & \myTEg{16}{1.55}{1.26}{7.84} & \myTE{7}{0.05}{0.02}{0.26} & \myTE{5}{0.89}{0.85}{1.03} & \myTE{12}{0.89}{0.44}{4.80} & \myTEg{15}{3.36}{1.96}{2.47} & \myTE{9}{0.45}{0.20}{1.21}
		\tabularnewline[0.05cm] 
		
		PSO & \myTE{16}{0.46}{0.04}{2.76} & \myTEg{15}{1.03}{0.46}{2.39} & \myTE{14}{0.86}{0.33}{1.38} & \myTEg{15}{1.05}{0.70}{2.42} & \myTE{13}{0.86}{0.48}{0.98} & \myTE{14}{0.60}{0.28}{0.77} & \myTE{11}{0.96}{0.87}{1.19} & \myTEg{14}{1.03}{0.53}{1.96} & \myTEg{6}{1.33}{1.21}{1.88} & \myTEb{5}{0.27}{0.12}{1.25}
		\tabularnewline[0.05cm] 
		
		PS & \myTEb{3}{0.01}{0.00}{0.04} & \myTEbb{1}{0.00}{0.00}{0.01} & \myTEb{2}{0.09}{0.01}{0.22} & \myTEb{2}{0.01}{0.01}{0.22} & \myTE{12}{0.84}{0.80}{0.85} & \myTE{13}{0.17}{0.05}{0.44} & \myTE{14}{0.97}{0.86}{0.99} & \myTE{2}{0.12}{0.04}{0.51} & \myTEbg{9}{1.47}{0.16}{2.79} & \myTEb{8}{0.38}{0.14}{1.34}
		\tabularnewline[0.05cm] 
		
		BADS & \myTEb{5}{0.02}{0.00}{0.08} & \myTE{3}{0.02}{0.00}{0.12} & \myTEbb{1}{0.06}{0.02}{0.20} & \myTEbb{1}{0.01}{0.00}{0.30} & \myTEb{11}{0.64}{0.12}{0.83} & \myTEbb{1}{0.01}{0.00}{0.02} & \myTE{10}{0.94}{0.85}{0.98} & \myTEbb{1}{0.04}{0.01}{0.18} & \myTEb{2}{0.90}{0.32}{1.38} & \myTEb{2}{0.17}{0.04}{0.36}
		\tabularnewline[0.05cm] 
		
		BayesOpt & \myTE{8}{0.06}{0.02}{0.35} & \myTE{8}{0.38}{0.13}{0.53} & \myTE{6}{0.20}{0.08}{0.42} & \myTEb{3}{0.22}{0.01}{1.08} & \myTE{7}{0.44}{0.32}{0.96} & \myTE{2}{0.03}{0.02}{0.06} & \myTEb{4}{0.88}{0.85}{0.98} & \myTE{3}{0.28}{0.16}{0.43} & \myTEg{7}{1.42}{1.03}{2.11} & \myTEb{6}{0.29}{0.11}{0.66} 
		\tabularnewline[0.05cm] 
		Fmincon & \myTEbb{1}{0.00}{0.00}{6.65} & \myTEg{17}{1.56}{2.59}{3.42} & \myTEg{17}{1.76}{0.38}{131} & \myTEg{17}{3.44}{0.81}{3.57} & \myTEg{17}{4.19}{1.31}{8.58} & \myTEg{17}{7.43}{2.89}{8.95} & \myTEg{17}{8.82}{0.86}{6.46} & \myTEg{17}{1.83}{5.23}{2.21} & \myTEg{13}{1.78}{1.19}{3.02} & \myTEg{17}{1.55}{0.38}{2.16}
		\tabularnewline[0.05cm]

		\bottomrule
	\end{tabular}
	\label{tab:allResults}
\end{table*}

\subsection{Benchmark Results} \label{sec:benchres}

Table \ref{tab:allResults} shows the results of the benchmark after a budget of $25 \, d$ evaluations. First, it is observed, that no optimizer clearly performs best for all test cases. Instead, the best performing algorithm (underlined) varies. However, CMAES, GA, PSO, and Fmincon (with the exception of test case 1), 
are consistently statistically significantly  
worse than the respective best performing algorithm and often worse than Rand.

Additionally, it is observed, that the respective best algorithm (underlined) significantly outperforms Rand with the exception of test case 9.  
Grid does not consistently outperform rand. At first glance BADS and PS seem to dominate the test cases with $d = 2$ (Cases 1. - 4.), whereas different variants of BO as well as BADS seem to dominate the remaining test cases. Therefore, these cases are analyzed separately below.

\subsubsection{Test Cases 1-4 (d = 2)} \label{sec:2dcases}

In Fig. \ref{fig:de2Benchmark.eps}, 
the average rank, average scaled median regret, as well as the average 80\% quantile of the regret are plotted as a function of the relative number of objective function evaluations. 
It is observed, that PS outperforms the other algorithms during the course of the optimization in all metrics starting from around $8 \, d$. Only when getting close to the maximum budget of $25 \, d$, BADS roughly ties PS. This observation agrees with Tab. \ref{tab:allResults}. The final scaled regret of BADS and PS are similar. Only for test case 2, the difference in scaled regret is statistically significant. 
As the best performing BO variant, MES-MA-V is identified. The other BO variants are not plotted for briefness. MES-MA-V and BayesOpt also perform substantially better than Rand, Grid, GA, PSO, CMAES and Fmincon.

\subsubsection{Test Cases 5-10 (d = 3-5)} \label{sec:moredcases}

For the higher dimensional test cases, first different acquisition functions and GP settings are compared. Afterwards the best performing BO variant is compared to the benchmark optimizers.

\paragraph{Acquisition Functions} \label{sec:ResMoreDaccFun}

Fig. \ref{fig:dg2Acc} shows, that MES and EI perform similarly if VDP for LCC is not used.  
Only UCB is worse when considering averaged median regret and averaged $80  \%$ quantile regret. 
It should be noted, that the important $\beta$ parameter of UCB (cf. Sec. \ref{sec:AccMdls}), which heavily influences explorativeness, was not fine tuned to the test cases.

\paragraph{GP Models}

In Fig. \ref{fig:dg2GPR}, the impact of 
different GP models is shown. It can be observed, that the proposed handling of crash constraints with virtual data points (MES-SE-V and MES-MA-V) yields an improvement over the versions without VDP (MES-SE-F and MES-MA-F). This is the case independent of whether the SE or MA kernel is used. Furthermore, MES-SE-V slightly outperforms MES-MA-V, with the difference increasing towards the maximum budget. In contrast to Sec. \ref{sec:ResMoreDaccFun}, using EI (EI-SE-V) over MES-SE-V slightly deteriorates performance. Using a quadratic trend function (MES-SEQ-V) seems to considerably deteriorate the performance of BO for the presented test cases. Finally, the usage of an arbitrarily chosen gamma hyper prior on the kernel length scales (MES-SEG-V) instead of the box hyper prior also leads to worse performance.

\paragraph{Comparison with Benchmark Optimizers}

Figure \ref{fig:dg2Benchmark} compares the best performing BO variant BO-MES-V with the benchmark optimizers. 
It is observed, that BADS performs best. However, BO-MES-V is competitive when the number of evaluations approaches $25d$. Fmincon, GA, CMAES, PSO, Grid and Rand are not competitive which agrees to Sec. \ref{sec:2dcases}. 
PS is the most competitive non-BO variant when considering average rank and average median scaled regret. The MATLAB reference implementation BayesOpt performs slightly worse than BO-MES-V, with the difference being biggest for the $80 \%$ quantile metric.

\subsection{Implications of Optimizer Choice on Controller Performance} \label{sec:TimeDomainAna}

In Sec. \ref{sec:benchres}, we focused on comparing different optimizers in terms of sample efficiency.  
As primary metric, the scaled regret was used. This metric does not take into account whether the gap between the overall best evaluation and the median random search regret is significant, 
 and as a result, whether the usage of an optimizer over random search comes with a practically relevant increase in closed-loop performance. 
 
 \vfill
 \noindent
 
\begin{minipage}{1.0\textwidth}
	 \strut\newline
	\centering
	\includegraphics[width = 0.7\textwidth]{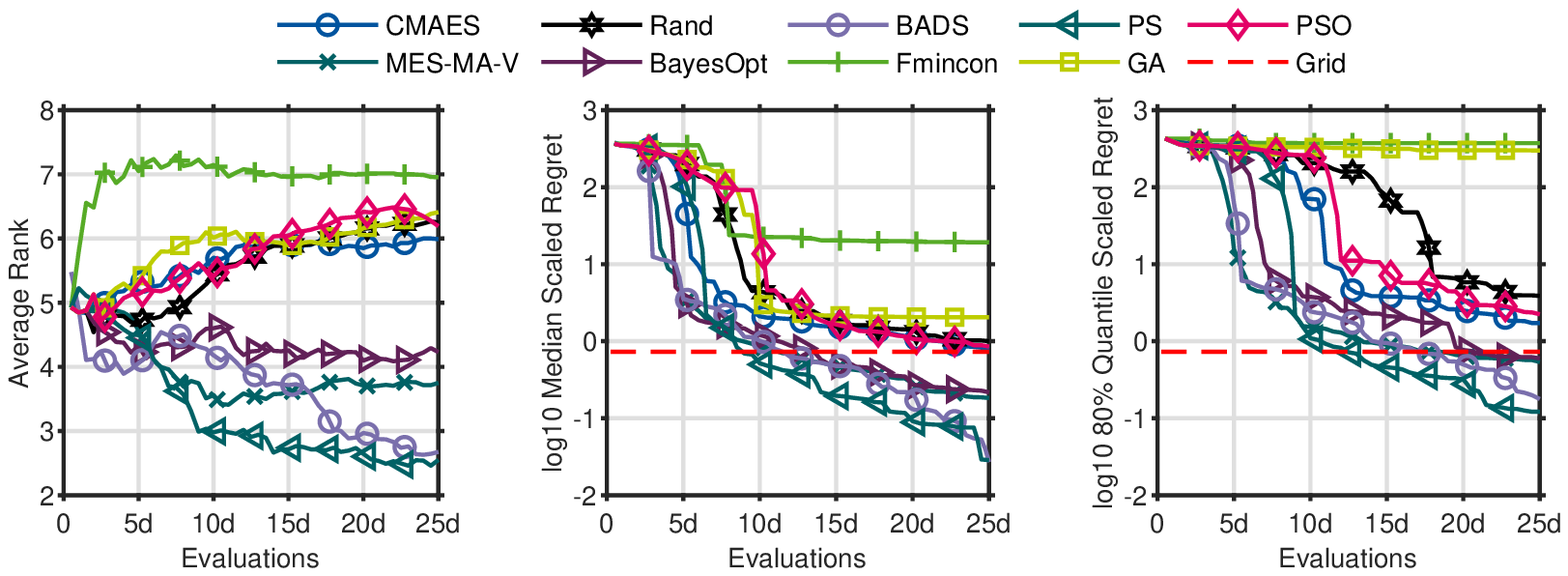}
	\captionof{figure}{Benchmark results for the two dimensional test cases 1 - 4. Pattern search (PS) and Bayesian adaptive direct search (BADS) perform best.}
	\label{fig:de2Benchmark.eps}
	\centering
	\includegraphics[width = 0.7\textwidth]{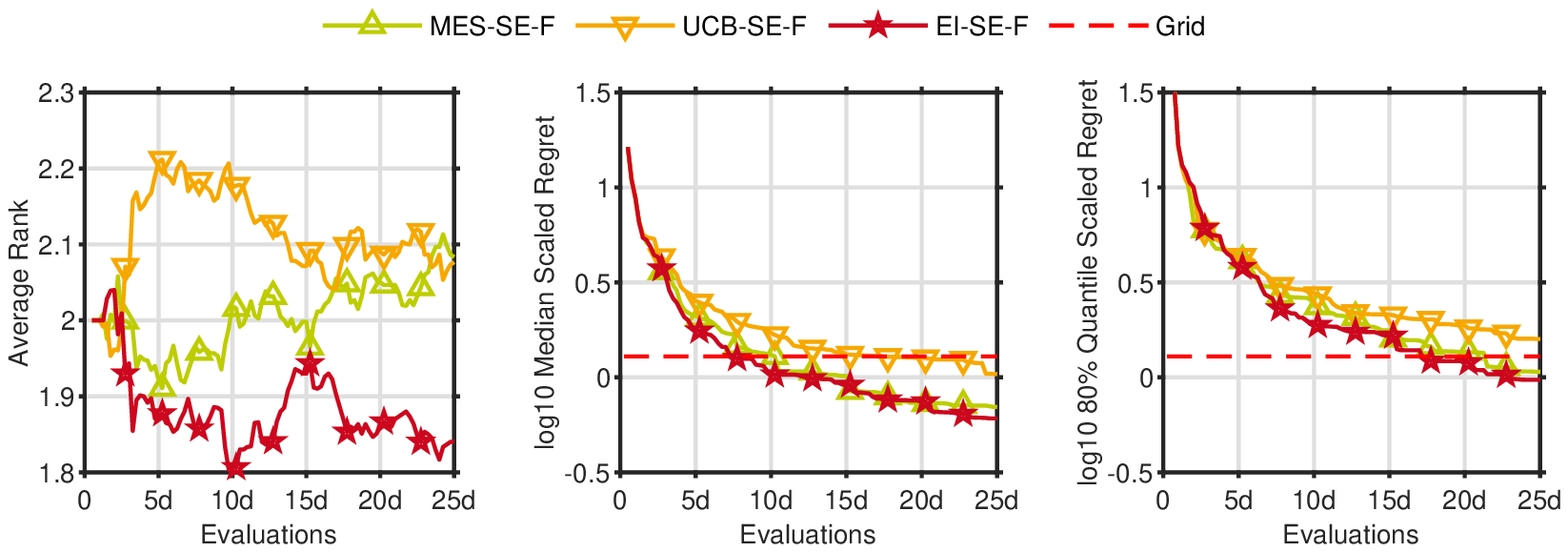}
	\caption{Benchmark results for different acquisition functions on the three to five dimensional test cases 5 - 10. Max-value entropy search (MES), and expected improvement (EI) perform similarly. UCB performs slightly worse.}
	\label{fig:dg2Acc}
	\centering
	\includegraphics[width = 0.7\textwidth]{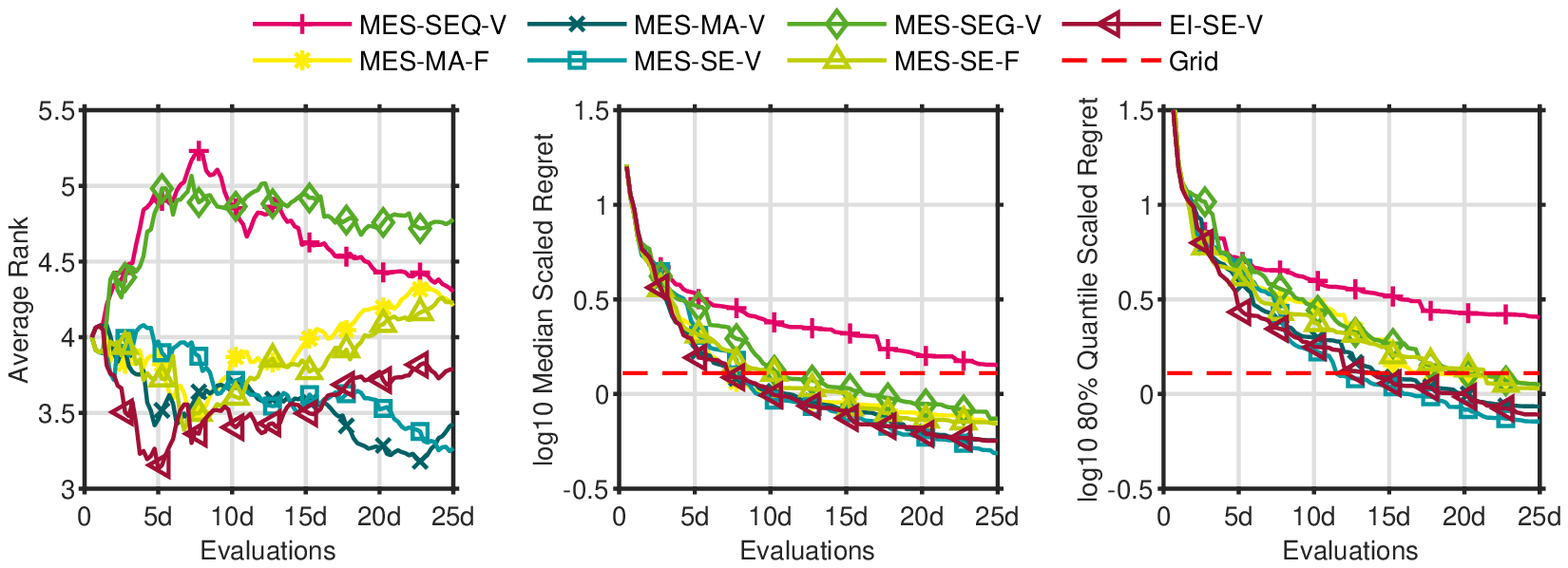}
	\caption{Benchmark results for different GP model settings on the three to five dimensional test cases 5 - 10. The BO variants with crash constraint handling using variable virtual data points (MES-SE-V, MES-MA-V and EI-SE-V) perform superior. Among them, the combination with Max-value entropy search and the squared exponential kernel (MES-SE-V) performs best. }
	\label{fig:dg2GPR}
	\centering
	\includegraphics[width = 0.7\textwidth]{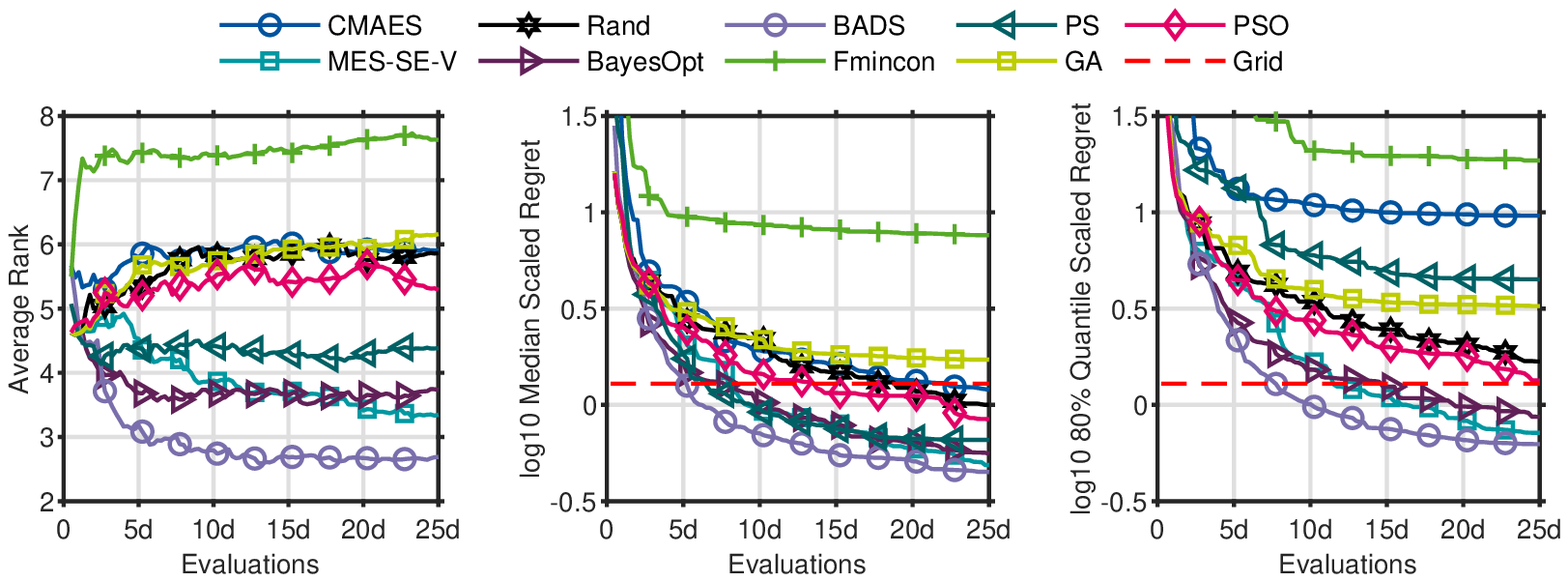}
	\caption{Benchmark results for the best performing BO variant with the benchmark algorithms on the three to five dimensional test cases 5 - 10. Bayesian Adaptive Direct Search (BADS) performs best with MES-SE-V becoming competitive when reaching $25 \, d$}
	\label{fig:dg2Benchmark}  
\end{minipage}

\newpage

\begin{table*}[h] 
	\caption{Performance increase achieved by using the recommended optimizers PS \& BADS over random search.}
	\centering
		\begin{tabular}{p{1.7cm}| >{\raggedleft}p{1.15cm} >{\raggedleft}p{1.15cm} >{\raggedleft}p{1.15cm} >{\raggedleft}p{1.15cm} >{\raggedleft}p{1.15cm} >{\raggedleft}p{1.15cm} >{\raggedleft}p{1.15cm} >{\raggedleft}p{1.15cm}   >{\raggedleft}p{1.15cm}   >{\raggedleft}p{1.3cm} }
			\toprule
			& \multicolumn{10}{c}{Test case No.}
			\tabularnewline
			&1 ($d = 2$)    &  2 ($d = 2$)          &  3 ($d = 2$)    & 4 ($d = 2$)  & 5 ($d = 3$)  & 6 ($d = 3$) &7 ($d = 4$) &8 ($d = 4$) &9 ($d = 4$) &10 ($d = 5$)
			\tabularnewline
			\midrule
			Random Search & $67.09$ ($100\%$)    &  $1.157$ ($100\%$) &  $9.58e-5$ ($100\%$)   & $0.224$ ($100\%$) & $42.14$ ($100\%$) & $0.669$ ($100\%$) & $65.79$ ($100\%$)  &$1.220$ ($100\%$) &$0.742$ ($100\%$) &$0.772$ ($100\%$)
			\tabularnewline[0.35cm]
			
			Recommended: PS + BADS & $65.50$ ($-2.4\%$)       &  $1.057$  ($-8.6\%$)       &  $9.08e-5$ ($-5.1\%$)   & $0.198$ ($-11.8\%$) & $41.47$ ($-1.6\%$) & $0.572$ ($-14.6\%$) & $65.14$ ($-1.0\%$) &$1.024 $ ($-16.1\%$) &$0.729$ ($-1.8\%$) &$0.751$ ($-2.8\%$)
			\tabularnewline[0.35cm]
			
			Overall best $J(\boldsymbol{ \theta}^*)$ & $65.48$  ($-2.4\%$)   &  $1.057$   ($-8.7\%$)      &  $9.04e-5$  ($-5.5\%$)  & $0.197$ ($-12.0\%$) & $40.27$ ($-4.4\%$) & $0.571$ ($-14.7\%$) & $55.77^*$ ($-15.2\%$) &$1.015$ ($-16.9\%$) &$0.609$ ($-18\%$) &$0.746$  ($-3.4\%$)
			\tabularnewline

			\tabularnewline
			\bottomrule
	\end{tabular}
	\label{tab:objFunVals}
\end{table*}
\FloatBarrier

Table \ref{tab:objFunVals} summarizes the median objective function values 
of 
random search and the recommended optimizers (PS for $d = 2$ and BADS for $d > 2 $). The overall best evaluation $J(\boldsymbol{ \theta}^*)$ is also given.
The usage of the recommended optimizer yields a decrease of $1\%$ to $16.1\%$ over random search with an average of $6.6 \%$. For example for test case 6, an improvement of $14.6 \%$ is reached.  This can be explained by it's very narrow global optimum (cf. Fig. \ref{fig:objFunAna}), which is unlikely to be found by random search. Interestingly, the performance increase over random search is not obviously dependent on $d$. 
Within the budget of $25 \, d$ the recommended optimizers are able to reach the global optimum with a margin of less than $1 \%$ for all test cases except 5, 7, and 9. 
For test case 5, the deviation is moderate at $2.8 \%$. For cases 7 and 9, the loss exceeds $10 \%$. This indicates that a budget of $25 \, d$ is not always sufficient. However, the optimal value found for test case 7, marked by (*), is located in an extremely narrow valley surrounded by infeasible evaluations (cf. Fig. \ref{fig:objFunAna} and Table \ref{tab:BenchTabel}). Most likely this parametrization would not be robust and therefore not desired for practical applications.

\begin{figure}[h]
	\centering
	\includegraphics[width=\columnwidth]{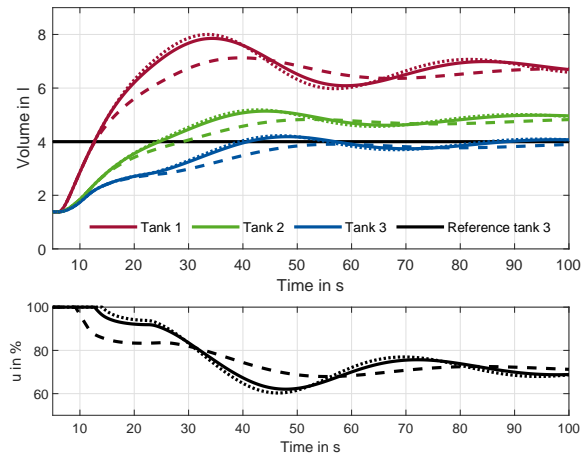}
	\caption{Comparison of the time-domain behavior for test case 3. \\ \textbf{Solid}: Optimized by pattern search (recommended for $d = 2$). \textbf{Dotted}: Overall best. \textbf{Dashed}: Optimized by random search.}
	\label{fig:Dreitankzeitbereich}
\end{figure}

In Fig. \ref{fig:Dreitankzeitbereich}, 
the impact of objective function value on time-domain behavior is visualized for test case 3. The reduction of the objective function value by $5.1 \%$ using PS over RS is clearly visible. The remaining difference of $0.4 \%$ between the recommended optimizer and the global optimum is barely noticeable.  

\section{Discussion} \label{sec:Dis} 
\subsection{Characteristics of the Benchmark Test Cases} \label{sec:DisChar}

From the analysis of the ten controller optimization problems, three main differences to most synthetic benchmarks are observed: possibility of deterministic noise, limited multi-modality, large areas with crashes (cf. Sec. \ref{sec:ObjFunAna}).  
The observed deterministic noise has been described before e.g. in the context of wing design \cite{Forrester.2006} and fault diagnosis \cite{Stenger.2019}. However, it is not present in simplified synthetic benchmarks e.g. the COCO benchmark \cite{NikolausHansen.2021}.  
Also the presence of a limited number of local minima is in stark contrast to for example the synthetic COCO benchmark \cite{NikolausHansen.2021}, where some objective functions have tens to hundreds local optima. The synthetic Hartmann and Michalewicz functions, which are used e.g. in \cite{Marco.2021} have equally many local optima as dimensions.  
Thirdly, it was shown that large fractions of the design parameter domain may lead to crashed simulations which is also mostly not considered in typical synthetic optimization benchmarks.

Therefore, results from synthetic benchmarks are not necessarily applicable to controller tuning problems. From the standpoint of BO research for control engineering, this suggests that algorithms may need to be tailored to these specific challenges. The low multimodality of the problems promote locally searching optimizers. Examples of locally searching BO variants include cautious BO \cite{Frohlich.2021}, GIBO \cite{SarahMuller.2021}, and BADS \cite{Acerbi.2017}.  
Alternatively convexity can be encoded in the GPR-model (cf. e.g. \cite{MarcoAlonso.2017,Brunzema.2022}). 
If deterministic noise is observed, it may be beneficial to optimize the noise hyperparameter of the GP and use a suitable infill criterion which prevents repetitive sampling (e.g. the reinterpolation procedure \cite{Forrester.2006, Stenger2019}) also in the case of deterministic simulative controller tuning. Additionally, crash constraints are an important issue also in simulative tuning. Methods to effectively deal with them have been proposed here or in \cite{Marco.2021}. As an alternative, safety constraints can be added in order to prevent sampling in unsafe regions. If these are available in analytical form they can directly be included \cite{Dorschel.2021}. If safe regions are approximated via GP models (e.g. \cite{Berkenkamp.14.02.2016}) sampling in unsafe regions is still possible and may still detoriate optimization performance \cite{Dorschel.2021}.

\subsection{Sample Efficiency of the Evaluated Algorithms}

It was confirmed that GA, CMAES, PSO, Fmincon, random search, and grid search lack in sample efficiency compared to BO using a budget of $25 \, d$. 
Furthermore, it was confirmed in agreement to \cite{Marco.2021} that effectively dealing with crash constraints is elementary for automated tuning in control engineering. 

However, surprisingly, pattern search (PS) performed best for the two dimensional test cases. As described in Sec. \ref{sec:relwork}, PS has not been considered much for controller tuning. One reason for its limited usage may be, that PS 
mainly searches locally. The convincing performance of PS therefore adds to the evidence presented in \hbox{Sec. \ref{sec:ObjFunAna}} that one dominating global optimum is present and the globalization capabilities of BO are not necessarily needed. 

For $d > 2$, BADS performs best.  
Therefore, results 
indicate that incorporating alternative search strategies is an effective way to deal with poor GP and a resulting poor BO performance. 
For the observed cases, poor GP performance may be caused by crashed evaluations and deterministic noise.  
Alternative hybrid BO algorithms were presented in \cite{RikkyR.P.R.Duivenvoorden.2017} (PSO) and \cite{Pitra.2016} (CMAES). If BADS is used in a multimodal environment, the usage of multistart strategies should be preferred over the increase of the budget in one run 
\cite{Acerbi.2017}. 

Although the best BO variant performed worse than PS or BADS respectively, the insights of the impact of BO design choices on sample efficiency are still valuable for more advanced problem formulations, where hybrid algorithms may not be usable in a straight forward manner. These settings include 
contextual BO (e.g. \cite{Fiducioso.6282019}) or multi-objective BO (e.g. \cite{GharibAli.2021}).    
In addition to the points made in Sec. \ref{sec:DisChar} w.r.t. crash constraints and weak multimodality, it was shown that the SE kernel can perform better than the MA kernel for $d > 2$ but not for $d = 2$. In \cite{LeRiche.2021}, it was also found that for some problem classes, the SE kernel can outperform the MA kernel. This is in contrast to the findings in \cite{Snoek.2012} where the MA kernel clearly outperformed the SE kernel. 
Additionally, results indicate that EI and and MES perform quiet similar on average and superior to UCB for $d > 2$. 
The ambiguity of the results, i.e. the dependency of the preferred BO set up on the test case  
support the findings of \cite{Turner.2021}, where BO ensembles performed best.  

Unlike to the results in \cite{LeRiche.2021}, the use of a quadratic mean function deteriorated sample efficiency in this contribution consistently. 

\section{Conclusion} \label{sec:con}

PS performed best for $d = 2$, and BADS performed best for $d > 2$. Therefore, these optimizers are recommended for single-objective controller optimization with crash constraints. Other metaheuristics such as GA, CMAES, and PSO as well as grid search, random search and Fmincon were shown to not be sample efficient. BO with crash constraint handling using virtual data points (VDP) was shown to be competitive to BADS for $d > 2$. Crash constraint handling was shown to be most critical to BO performance. 

The usage of the recommended algorithms enables optimization based tuning in simulation as a reliable tool also for small budgets (here $25 \, d$) and therefore high-fidelity simulation environments. This way, hand tuned parametrizations can be validated or improved upon, and fair comparisons between different controller or filter structures can be made without substantial manual effort. 

Controller optimization problems pose specific challenges for optimizers, which are not typically found in synthetic benchmarks. As a result, optimization algorithms may need to be specifically tailored for controller tuning.   
Additionally, this contribution motivates the development and application of hybrid methods such as BADS, 
locally searching BO variants, and ensemble BO for advanced problems in controller tuning such as multi-objective or contextual optimization.  

\bibliographystyle{IEEEtran}

\bibliography{IEEEabrv,Bench}

\end{document}